\documentclass[prl,twocolumn,showpacs,epsfig,longeq]{revtex4}
\usepackage{graphicx}
\usepackage{dcolumn}
\usepackage{amssymb}
\usepackage{bm}
\usepackage{epsfig}
\usepackage{psfrag}
\usepackage[dvips]{color}
\usepackage[utf8x]{inputenc}
\usepackage{ucs}
\usepackage{amsmath}
\usepackage{subfig}
\usepackage{amsfonts}
\usepackage{amssymb}
\usepackage{graphicx}
\usepackage{dcolumn}
\usepackage{bm}
\usepackage{latexsym}
\usepackage{hyperref}
\newcommand{\bsf}[1]{\textsf{\textbf{#1}}}

\def\3dots{\:\raisebox{-0.5ex}{$\stackrel{\textstyle.}{:}$}\:}
\def\beq{\begin{equation}}
\def\eeq{\end{equation}}
\def\bea{\begin{eqnarray}}
\def\eea{\end{eqnarray}}
\newcommand{\look}[1]{\textcolor{black}{#1}}
\newcommand{\lookk}[1]{\textcolor{black}{#1}}

\begin{document}

\title{Anisotropic Isometric Fluctuation Relations in experiment
and theory on a self-propelled rod}
\author{Nitin Kumar$^1$, Harsh Soni$^{1,2}$, Sriram Ramaswamy$^{2,*}$ and A.K.
Sood$^1$}
\affiliation{$^1$Department of Physics, Indian Institute of Science, Bangalore
560 012, India}
\affiliation{$^2$TIFR Centre for Interdisciplinary Sciences, Tata
Institute of Fundamental Research, 21 Brundavan Colony, Osman Sagar
Road, Narsingi, Hyderabad 500 075, India}
\altaffiliation{On leave from the Department of Physics, Indian Institute
of Science, Bangalore}
\date{\today}
\begin{abstract}
The Isometric Fluctuation Relation (IFR) [P.I. Hurtado et al., PNAS \textbf{108}, 7704 (2011)] relates the relative probability of current fluctuations of fixed
magnitude in different spatial directions. We test its validity in an experiment
on a tapered rod, rendered motile by vertical vibration and immersed in a sea
of spherical beads. We analyse the statistics of the velocity vector of the rod
and show that they depart significantly from the IFR of Hurtado et al.
\lookk{Aided by a Langevin-equation model we show that our measurements are
largely described by an anisotropic generalization of the IFR [R. Villavicencio
et al., EPL \textbf{105}, 30009 (2014)], with no fitting parameters, but with a
discrepancy in the prefactor whose origin may lie in the detailed statistics of
the microscopic noise.} The experimentally determined
Large-Deviation Function of the velocity vector has \look{a kink} on a curve in
the plane.

\end{abstract}
\maketitle

Fluctuation relations (FR) go beyond the second law of thermodynamics by
quantifying the relative probability of short-time, small-scale
entropy-consuming events \cite{reviews}. Experimental evidence for their
validity is surprisingly widespread
\cite{ciliberto1,Goldburg,ciliberto2,Douarche}, even in non-thermal noisy
systems nominally outside their realm of applicability
\cite{ciliberto3,menon,puglisi,asood,sayantan,saroj}. A more stringent symmetry property of
thermal systems, the Isometric Fluctuation Relation (IFR) \cite{PNASIFR}, has
been derived recently for isotropic systems, describing the relative
probabilities of observing currents of equal magnitude in different directions,
not necessarily diametrically opposed as in the standard FR \cite{reviews}.

\look{In this Rapid Communication} we present the first experimental \look{observations of behaviour consistent with} the IFR, on a
{macroscopic, fore-aft asymmetric rod which executes self-propelled
\cite{sano,vj,Kudrolli2008,NK,NatComm,timerev} motion through a background of non-motile
spheres.} Our main results are as follows: (i) Our
\textit{anisotropic} experimental system deviates substantially from the
predictions of \cite{PNASIFR}. (ii) We show that the symmetry properties of the
\look{Large-Deviation (LD)} function correspond to the Anisotropic IFR of
\cite{touchette2}, with which we are able to show a parameter-free agreement
through a simplified description based on an ansotropic single-particle Langevin
equation. \lookk{There remains a 20 \% discrepancy in the prefactor whose
origin,
we speculate, could lie in non-Gaussianity in the microscopic noise, not
accounted for in our Langevin model.} (iii) We find that the measured LD
Function of the
velocity vector exhibits a kink where the velocity component along the rod
axis vanishes. \look{We speculate why these behaviours should arise in an apparently non-time-reversible system.}

\begin{figure}
\centerline{\includegraphics[width=0.5\textwidth]{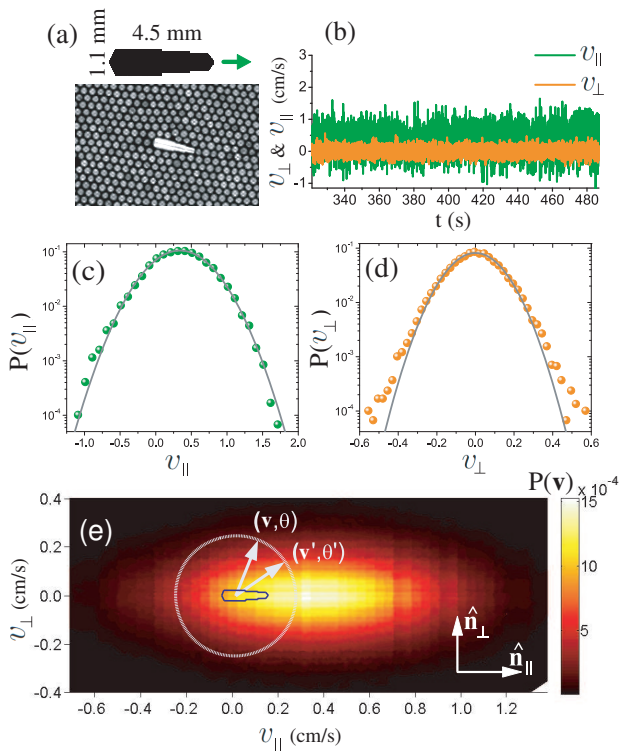}}
\caption{(a) Geometry of the polar particle: the thick arrow along the particle's axis indicates the mean direction of ``self-propelled'' motion and typical experimental screen-shot depicting the particle moving its way through the crystalline medium consisting of 0.8 mm aluminium beads in a confined two-dimensional cell. (b) Time-series of velocity components $v_{\parallel}$ $\&$ $v_{\perp}$ of the particle. Positive direction is along the green arrow in (a). Probability distributions of $v_{\parallel}$, (c) and $v_{\perp}$, (d) with solid lines showing a Gaussian fit. (e) Resulting probability distribution of velocity $\mathbf{v}\equiv v_{\parallel}\hat{\mathbf{n_{\parallel}}} + v_{\perp}\hat{\mathbf{n_{\perp}}}$  of the particle. Circle shows a constant-velocity contour and the arrows representing isometric vectors subtending angles $\theta$ $\&$ $\theta'$ with the horizontal axis.}
\label{setup}
\end{figure}

In our experimental setup \cite{NK} a single geometrically polar
brass rod, 4.5 mm long with diameter 1.1 mm at the thick end is placed
amidst a monolayer of spherical aluminium beads of diameter 0.8 mm (Fig. \ref{setup}(a)). The beads lie on an circular aluminium plate 13 cm in diameter, covered by a glass
lid at 1.2 mm above the surface, thus forming a confined two-dimensional system.
The bead area fraction, based on their projected images, is 0.83. A
permanent-magnet shaker (LDS V406-PA 100E) drives the plate
sinusoidally in the vertical direction with amplitude $a_{0}$ and frequency $f$
= 200 Hz, corresponding to dimensionless shaking strength $\Gamma \equiv
{a_{0}(2 \pi f)^{2}}/{g}= 6.5$, where $g$ is the acceleration due to gravity.
The rod transduces the vibration into predominantly forward motion as
indicated by the arrow in Fig. \ref{setup}(a) \cite{NK,sano,vj}. The bead
medium is both athermal noise source and obstacle course for the motion
of the polar rod (see Supplementary Video). A high-speed camera (Redlake
MotionPro X3) records the dynamics of the particle at a rate of 50 frames per
second. A typical experimental snapshot is shown in Fig. \ref{setup}(a). The
images were analysed in ImageJ \cite{ImageJ} to calculate {the
instantaneous in-plane position ${\bf R}(t)$ and orientation, i.e., the unit
vector $\hat{\mathbf{n_{\parallel}}}(t)$ from the thick to the thin end of the
rod  of the polar rod, and its two-dimensional velocity
vector ${\bf v}(t)$ at time $t$ \textit{defined} as the discrete time
derivative of ${\bf R}(t)$ between successive frames. Note that
${\bf v}(t)$ is already coarse-grained in time with respect to the true
microscopic velocity. The plate vibrates at 4 times the frame rate, and the
collisions of the rod with plate, lid and beads take place at irregular
instants.}

{We resolve ${\bf v}(t)$ into components $(v_{\parallel}(t),v_{\perp}(t)) =({\bf
v}(t)\cdot\hat{\mathbf{n_{\parallel}}}(t),{\bf
v}(t)\cdot\hat{\mathbf{n_{\perp}}}(t))$ along and transverse to
$\hat{\mathbf{n_{\parallel}}}(t)$ [Fig. \ref{setup}(e)].
In earlier work \cite{NK} on this system we studied the statistics of $v_{\parallel}$
alone and found a large-deviation function with a kink at zero as predicted in
several theoretical treatments of forced Brownian motion amidst periodic
obstacles \cite{Seifert,ASEP,MolecularMotor,Budini}. The results of \cite{NK}
encourage us to further probe its nonequilibrium fluctuations and look for a
possible IFR. Fig. \ref{setup}(b) shows the instantaneous time-series of $v_{\parallel}$,
whose nonzero mean signals systematic ``self-propulsion'', and $v_{\perp}$, with mean
zero. The statistical anisotropy of the dynamics is clear from the probability
distributions of $v_{\parallel}$ and $v_{\perp}$ in Figs. \ref{setup}(c) and (d)
respectively, which show a much greater dispersion along the rod than transverse
to it \cite{footnote_temperature}. Fig. \ref{setup}(e) shows the distribution
$P(\mathbf{v})$ of the two-dimensional velocity vector, peaked at a non-zero
$v_{\parallel}$ with significant weight in all directions including backwards. Our
experiments can thus explore the applicability of the IFR in a much larger
angular range than in the numerical study of \cite{PNASIFR}.}
Before presenting our experimental findings in detail, we build a
minimal single-particle Langevin equation model for the dynamics of the
two-dimensional position ${\bf R}(t) = (X(t), Y(t))$ of the polar rod as a
function of time $t$, ignoring inertia. This will allow an independent
determination of parameters required later in the paper, and help relate our findings to the anisotropic \cite{touchette2} isometric fluctuation relation in a simplifying limit. As we work in a frame
fixed in the particle, with $X$ and $Y$ defined with respect to the particle
orientation, the dynamics of ${\bf n}_\|(t)$ does not enter our analysis. The
governing Langevin equations read
\begin{equation}\label{compacteom}
\bm{\Gamma} \cdot \dot{\bf R} = {\bf F} + \bsf{N} \cdot {\bf f}(t)
\end{equation}
where ${\bf F}$, $\bm{\Gamma}$ are the systematic force and damping matrix.
${\bf F}$ gets contributions from the propulsive force driving the particle
along ${\bf n}_\|(t)$ as well as from the spatial structure of the obstruction
to motion posed by the bead medium. The random kicks that the rod receives from
beads and plate are described by a noise ${\bf f}$. The physical noise
covariance is $\bsf{N} \cdot \bsf{N} = 2 \bsf{D} \bm{\Gamma} \cdot \bm{\Gamma}$,
which defines the diffusion tensor $\bsf{D}$. We assume the
unit-strength noise ${\bf f}$ to be Gaussian and white, a plausible assumption
whose validity can be tested only by comparison to experiment.
Our interest here, as in \cite{PNASIFR}, is in large deviations of the
macroscopic current $\int_{\bf r}{\bf J}({\bf r},t)$.  For the present
single-particle system the current density ${\bf J}({\bf
r},t)= \delta({\bf r} - {\bf R}(t)){\bf v}(t)$ at point ${\bf r}$ at time
$t$, where ${\bf v} \equiv \dot{\bf R}$, so that the macroscopic current is
simply ${\bf v}(t)$.

We would like to extract the parameters in \eqref{compacteom} from the
experiment. We
define ${\bf V}_\tau(t)=\tau^{-1}\int^{t+\tau}_t {\bf v}(t') dt'$, the current
coarse-grained on a timescale $\tau$. For the smallest accessible $\tau$, of
order the inverse frame rate, it is reasonable to suppose that the structure of
the bead medium does not affect the dynamics of the rod significantly. We can
then assume ${\bf F}$ in \eqref{compacteom} depends only on time.
Define $\bm{\Pi} \equiv \bsf{N}^{-1} \bm{\Gamma}$, ${\bf S} \equiv \bsf{N}^{-1}
{\bf F}$, and $\bm{\mathcal{S}}_\tau(t)=
\tau^{-1} \int^{t+\tau}_{t}{\bf S}(t')dt' $, we can
construct, from \eqref{compacteom} and
the statistics of the noise sources, the probability
density
\begin{equation}
\label{probVtau}
P_{\tau}(\textbf{V}_{\tau} = {\bf V})=
\mbox{det}\bm{\Pi} \left({\tau \over 2 \pi}\right)^{d/2}
\exp \left[- {\tau \over 2} (\bm{\Pi} \cdot {\bf V} -
\bm{\mathcal{S}}_\tau)^2 \right]
\end{equation}
for ${\bf V}_\tau(t)$ to take a value
${\bf V} = (V_{\parallel},V_{\perp})$, via its
moment-generating function as shown in Appendix.
If we approximate $\bm{\mathcal{S}}_\tau(t)$ by $\bm{\Pi} \cdot {\bf v}_0$,
where ${\bf v}_0$ is the steady-state average velocity, which would be exact for
$\tau \to \infty$ and is reasonable for the present case where the inverse of frame rate, $\tau_{f}$,
is \look{four times the oscillation period},
(\ref{probVtau}) becomes
\begin{equation}
\label{probVtau_av}
P_{\tau}(\textbf{V}_{\tau} = {\bf V})=
\left(\mbox{det}\bsf{D}\right)^{-1/2} \left({\tau \over 2 \pi}\right)^{d/2}
e^{- {\tau \over 4} ({\bf V} -  {\bf v}_0)^T \bsf{D}^{-1}({\bf V} -  {\bf
v}_0)}
\end{equation}
where the inverse diffusion tensor $(2\bsf{D})^{-1} = \bm{\Pi}^T \bm{\Pi}$
can be seen to provide a natural inner product.
Below we use the form (\ref{probVtau_av}) to extract values for $v_0$
and the diffusivities from our data. Eq. (\ref{probVtau}), the
result of ignoring the position dependence of the forcing in \eqref{compacteom},
trivially obeys an anisotropic IFR \cite{touchette2} because its
large-deviation function, from \eqref{probVtau_av}, is quadratic: 
two coarse-grained currents
$\textbf{V}$ and $\textbf{V}'$ satisfying 
\begin{equation}\label{ifr1}
{\bf V}^T\bsf{D}^{-1} {\bf V} = {\bf V}'^T\bsf{D}^{-1} {\bf V}'
\end{equation}
i.e., which lie on the ellipse ${\bf V}^T\bsf{D}^{-1} {\bf V} =
\text{constant}$, obey
\begin{equation}\label{ifr}
\lim_{\tau \to \infty }\frac{1}{\tau} \ln
\dfrac{P_{\tau}(\textbf{V}_\tau = {\bf V})}{P_{\tau}(\textbf{V}_\tau =
{\bf V}')}=\bm{ \epsilon}
\cdot (\textbf{V}-\textbf{V}'). 
\end{equation}
with
\begin{equation} \label{epsilon}
\bm{ \epsilon} = {\bf v}_0^T (2\bsf{D})^{-1}.
\end{equation}
\lookk{In the event that $D\Gamma$ turns out to be proportional to the unit tensor, with coefficient $T_{eff}$,} local detailed balance holds. Then $\bm{\epsilon} = \bm{\Gamma} {\bf
v}_0 /2 T_{eff}$ is the drag force scaled by effective temperature, and
\eqref{ifr} becomes a true fluctuation relation for the \textit{power}. 

For comparison with our experiments, let us consider currents ${\bf V}$,
${\bf V}'$ with \textit{equal} magnitude $V$. The result of
\cite{PNASIFR} can then be
re-expressed, for the case of a diagonal $\bsf{D}
= \mbox{diag}(D_{\parallel}, D_{\perp})$, as 

\begin{eqnarray}\label{new1}
\nonumber
\lim_{\tau \to \infty }&\dfrac{1}{\tau}\ln
\dfrac{P_{\tau}(\textbf{V}_\tau ={\bf V})}{P_{\tau}(\textbf{V}_\tau = {\bf
V}')}
= V \left[ \dfrac{v^{\parallel}_0}{2 D_{\parallel}} (\cos \theta - \cos
\theta') \right.\\
\nonumber
&+ \left. \dfrac{v^{\perp}_0}{2 D_{\perp}} (\sin \theta - \sin \theta')\right]\nonumber
\\
&+\dfrac{V^2
(D_{\parallel}-D_{\perp})}{4D_{\perp} D_{\parallel}}( \cos^2 \theta -\cos^2 \theta')
\end{eqnarray}
Although the large-deviation function for our system will not have
the quadratic form
implied by \eqref{probVtau_av}, the foregoing calculation gives us a
value [Eq. \eqref{epsilon}] for $\bm{ \epsilon}$ in \eqref{ifr} in terms of
independently measurable quantities. This allows a parameter-free comparison of
our measurements to the Anisotropic Fluctuation Relation (AIFR) \cite{touchette2} in
the form \eqref{ifr} or \eqref{new1}. 
\begin{figure}[!t]
\centerline{\includegraphics[width=0.5\textwidth]{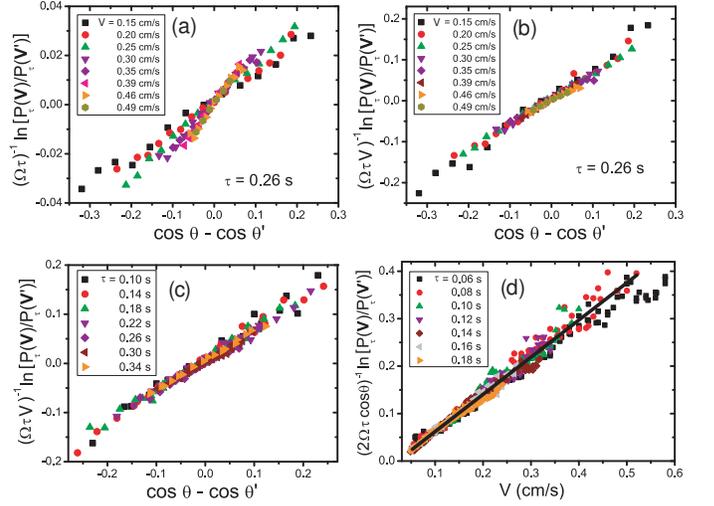}}
\caption{(a) A typical plot of $(\Omega\tau)^{-1}$ ln $[P_{\tau}(\mathbf{V})/P_{\tau}(\mathbf{V'})]$ vs. $\cos \theta$ - $\cos \theta'$ over various constant-velocity contours for $\tau$ = 0.26 s showing a linear trend for all $V$. (b) Data scaling of $(\Omega\tau V)^{-1}$ ln $[P_{\tau}(\mathbf{V})/P_{\tau}(\mathbf{V'})]$ vs. $\cos \theta$ - $\cos \theta'$. (c) Scaling of $(\Omega\tau V)^{-1}$ ln $[P_{\tau}(\mathbf{V})/P_{\tau}(\mathbf{V'})]$ with $\tau$ variation. Here each $\tau$ line contains all the $V$ values as in (b). (d) Plot of (2$\Omega\tau$cos$\theta$)$^{-1}$ ln $[P_{\tau}(\mathbf{V})/P_{\tau}(\mathbf{V'})]$ vs $V$ for various $\tau$ for the special case when $\theta - \theta' = 180^{0}$. Here $\theta$ varies between $-30^{0}$ to $30^{0}$ in the step of $10^{0}$  for all $\tau$.}
\label{valid}
\end{figure} 

We return now to the experiments.
The polar rod is propelled only along its nose, i.e., in the
$x$ direction. Thus y-component of $\bf{F}$ in \eqref{compacteom} is zero, so that only the component $v_0^{\parallel}$ of
the mean velocity is nonzero.
From the measured time series of positions we can extract the distribution of
${\bf V}_{\tau}$ for the shortest time accessible ($\tau_f$) and infer $v_0^{\parallel}$, $D_{\parallel}$ and $D_{\perp}$ by fitting the
measured distribution to (\ref{probVtau_av}). We obtain $D_{\parallel}$ and
$D_{\perp}$ as $\left\langle (V_{\parallel}(t) - v_0)^2\right\rangle_t \tau)/2$ and
$(\left\langle (V_{\perp}(t))^2\right\rangle_t \tau)/2$ respectively where $\tau$ = $\tau_f$ =
0.02 s and $v_0$ = 0.34 cm s$^{-1}$. We now plot constant velocity
contours centered at ($V_{\parallel}$ = 0, $V_{\perp}$ =
0) and consider sets of $\mathbf{V_{\tau}}$ and $\mathbf{V_{\tau}'}$
at angles $\theta$ and $\theta'$ with respect to
$\hat{\mathbf{n_{\parallel}}}$ (Fig. \ref{setup}(e)). We consider overlapping
azimuthal bins
of $\mathbf{V_{\tau}}$ in order to improve statistics, and obtain the
probability density $P(\mathbf{V_{\tau}})$. Defining $\Omega(\theta, \theta') =
\dfrac{v^{\parallel}_0}{2 D_{\parallel}} + \dfrac{V (D_{\parallel}-D_{\perp})}{4D_{\perp} D_{\parallel}}( \cos \theta +
\cos\theta')$, we plot
$(\Omega\tau)^{-1}$ln$\left[P_{\tau}(\mathbf{V})/P_{\tau}(\mathbf{V}')\right]$
as a function of $\cos \theta$ - $\cos \theta'$, as shown in Fig. \ref{valid}(a)
for a typical $\tau$ = 0.26 s at various values of $V$.
The trend is linear for all $V$ in agreement with Eq. \ref{new1} (please recall that $v_0^{\perp}$ = 0). Furthermore,
a plot of $(\Omega \tau
V)^{-1}$ln$\left[P_{\tau}(\mathbf{V})/P_{\tau}(\mathbf{V}')\right]$ shows a
clear scaling with $V$ for  $\tau$ = 0.26 s in Fig. \ref{valid}(b).  Similar
data collapse is seen for all $\tau>0.10$ s. Now we plot $(\Omega \tau
V)^{-1}$ln$\left[P_{\tau}(\mathbf{V})/P_{\tau}(\mathbf{V}')\right]$ for various
$\tau$ in Fig. \ref{valid}(c). Here each $\tau$ line contains various $V$ values
as in Fig. \ref{valid}(b). Overlapping lines for all $\tau$ confirm the $\tau$
scaling of the AIFR. From Eq. \ref{new1}, the expected slope in Fig.
\ref{valid}(b) or (c) shold be unity. Our experiment, using the estimates for
$v_0$, $D_{\parallel}$ and $D_{\perp}$, finds a number close to 0.8. Perhaps this reflects
limitations inherent in using estimates of mean speed and diffusivity from
short-time data. A possible reason for this discrepancy is that the noise is not
Gaussian \look{\cite{footnote_nongaussian}}: Fig. \ref{setup}(d) shows significant non-quadratic tails in the
logarithm of the distribution of the small-$\tau$ averaged velocity, giving some
credence to this suggestion.

As a check, we consider the special case of oppositely directed vectors, i.e.
$\mathbf{V}'$=$-\mathbf{V}$, (implying $\theta-\theta'$ =
$180^{0}$). Fig. \ref{valid}(d) shows the plot of $(2\Omega\tau cos\theta)^{-1}$ln$[P_{\tau}(\mathbf{V})/P_{\tau}(\mathbf{V}')]$ as a function of $V$ for various $\tau$. Here $\theta\in(-30^{0}, 30^{0})$
separated by a step of $10^{\circ}$. A clean collapse of data is observed with
the same slope as in Fig. \ref{valid}(c). This illustrates compliance with
the standard FR \cite{reviews} as a special case of AIFR.

\begin{figure}[!t]
\centerline{\includegraphics[width=0.5\textwidth]{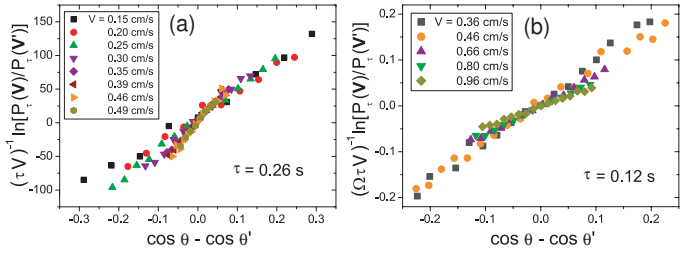}}
\caption{(a) Ignoring the anistropy term in Eq. \ref{new1}, i.e. assuming $D_{\parallel} = D_{\perp}$, no scaling is observed for $(\tau V)^{-1}$ ln $[P_{\tau}(\mathbf{V})/P_{\tau}(\mathbf{V'})]$ vs. $\cos \theta$ - $\cos \theta'$. (b) AIFR analysis for the particle moving on a bare substrate in the absence of bead medium. No collapse of data for $(\Omega \tau V)^{-1}$ ln $[P_{\tau}(\mathbf{V})/P_{\tau}(\mathbf{V'})]$ vs. $\cos \theta$ - $\cos \theta'$ suggesting the presence of the noisy medium for AIFR to hold.}
\label{fail}
\end{figure}

For completeness, we analyse the velocity fluctuations in the experiment in
terms of the IFR for isotropic systems \cite{PNASIFR}.
It can be obtained by setting $D_{\parallel}$=$D_{\perp}$ in Eq. \ref{new1} which
gives $\Omega=v_0/2D_{\parallel}$ independent of $\theta, \theta'$. We plot $(\tau
V)^{-1}$ln$\left[P_{\tau}(\mathbf{V})/P_{\tau}(\mathbf{V}')\right]$ against 
$\cos \theta$ - $\cos \theta'$ for $\tau$ = 0.26 s. Fig. \ref{fail}(a) shows
that the isotropic IFR is not satisfied. We check that this trend is
seen for all $\tau$. We find a trend of increasing slope with $V$ which cannot
be explained without the inclusion of the second term in $\Omega$. This
reiterates, experimentally, the importance of anisotropy as pointed out by 
\cite{touchette2}. We check in addition the role of the noise and hindrance to
motion provided by the bead medium, by examining the velocity
fluctuations of the particle for its motion on a bare substrate in the absence
of any medium. Now the source of all the noise is multiple random collisions of
the particle with the base and the lid. After an analysis along the same lines
as above, we plot in Fig. \ref{fail}(b) $(\Omega\tau
V)^{-1}$ln$\left[P_{\tau}(\mathbf{V})/P_{\tau}(\mathbf{V}')\right]$ as function
of $\cos \theta$ - $\cos \theta'$ for $\tau = 0.12 s$. Interestingly, we find a
significant failure of data-collapse in this case. Similar behaviour is
observed for all $\tau$. The presence of the bead medium appears to be
important for the AIFR to hold, a result consistent with our earlier findings \cite{NK}.

It is curious that fluctuation relations, which are a derived consequence of
microscopic time-reversibility, should arise in a system with a unidirectional
flow of energy. It would appear that in the present experiment the only
significant effect of this energy input is its transduction into directed motion
by the polar rod. The central role of the bead medium is also a puzzle. We
speculate the medium provides closely related processes governing the damping
and diffusion matrices $\bm{\Gamma}$ and $\bsf{D}$ in \eqref{compacteom} of the
polar rod, thus giving rise to an effective local detailed balance. In addition,
perhaps the multiple collisions of the polar particle with the bead medium
provides suppress correlated movements of the rod due to rolling or sliding, 
which may be present in motion on a bare substrate.

\begin{figure}
\centerline{\includegraphics[width=0.5\textwidth]{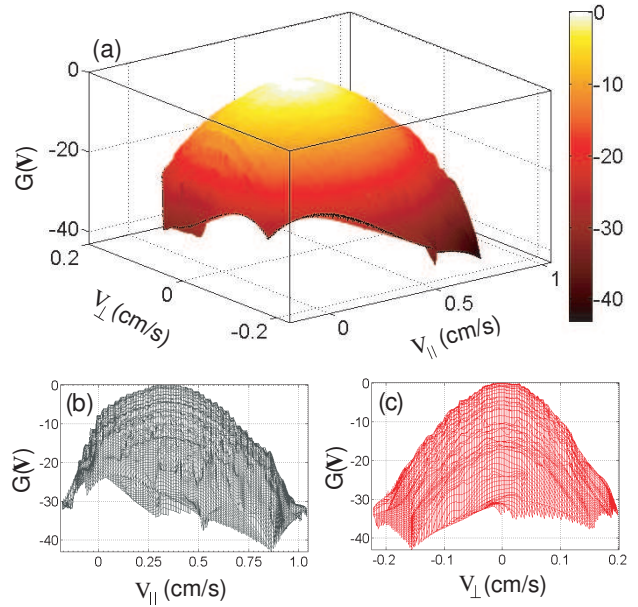}}
\caption{(a) The Large Deviation Function for the velocity fluctuations \look{for the case of the polar rod moving through the bead medium is non-paraboloid}. (b) The side-view along the $V_{\perp}$-axis shows an abrupt fall which resembles the kink-like feature observed in \cite{NK}. (c) The side-view along the $V_{\parallel}$-axis.}
\label{LDF}
\end{figure}

Lastly, we extract the Large Deviation Function (LDF) \look{for the velocity vector of a polar rod moving through the bead medium}, $G(\mathbf{V})$ defined
as $P_{\tau}(\mathbf{V}) = A_{\tau}$exp$(\tau G(\mathbf{V}))$ ($G(\mathbf{V})$
$<$ 0) in the limit $\tau \rightarrow \infty$ \cite{touchette}. Here we take
$A_{\tau} = max[P_{\tau}(\mathbf{\mathbf{{V}}})] = P_{\tau}(\langle
\mathbf{V}\rangle=v_0)$ which is independent of $\mathbf{V}$ as
required. We plot the asymptotic LDF as shown in Fig. \ref{LDF}(a). Clearly the shape of LDF is far from paraboloid. There is a sharp
drop in $G(\mathbf{V})$ below $V_{\parallel} = 0$ plane which is clearer when viewed
along the $V_{\perp}$-axis (Fig. \ref{LDF}(b)). This is related to the kink observed
in earlier work \cite{NK}. Fig. \ref{LDF}(c) shows that the LDF is parabolic
when viewed along the $V_{\perp}$ axis.

We point out here that earlier tests of the IFR, e.g., energy diffusion
on a two-dimensional lattice, a hard-disk fluid in a temperature gradient
\cite{PNASIFR} were numerical. Moreover, these
simulations saw no negative events and fluctuations were only \textit{along} the
imposed gradient, resulting in a test of the IFR in a very limited range of
angles $\theta$. Our experimental study, with substantial noise, allows an
exploration over a large $\theta$ range.

To summarise: we find detailed experimental support for the anisotropic variant
\cite{touchette2} of the Isometric Fluctuation Relation \cite{PNASIFR}, for a
self-propelled granular particle moving through a monolayer of spherical
beads. \lookk{We suggest that the discrepancy between the observed and
predicted prefactor arises from non-Gaussianity in the form of heavy tails in
the microscopic noise, not accounted for in our Langevin equation model. More
work is needed to address this.} Our
measurements are consistent with an earlier study of a Fluctuation
Relation, and include an extension of our earlier
observation of a kink in the Large-Deviation Function of the velocity vector of
the particle. Our treatment through an anisotropic Langevin equation
sheds some light on the circumstances in which such relations should arise, and allows a
parameter-free test of the theory. 

For support, NK thanks the University Grants Commission, India, HS thanks the
CSIR, India, and AKS and SR acknowledge a J C Bose Fellowship of the DST, India.

\appendix

\section{APPENDIX: Detailed calculation for Anisotropic isometric Fluctuation Relation}\label{apend1}

The overdamped Langevin equations for a particle subjected to a force $\textbf{F}(t)$ in $d$ dimensions can be written as  
\begin{equation}\label{compacteomsup}
\bm{\Gamma} \cdot \dot{\bf R} = {\bf F} + \bsf{N} \cdot {\bf f}(t).
\end{equation}
Multiplying above Eq. by $\bsf{N}^{-1}$ gives

\begin{equation}\label{compacteomsup}
\bm{\Pi} \cdot \dot{\bf R} = {\bf S} + {\bf f}(t),
\end{equation}
where $\bm{\Pi} \equiv \bsf{N}^{-1} \bm{\Gamma}$, ${\bf S} \equiv \bsf{N}^{-1}
{\bf F}$.
The current for the single particle at point ${\bf r}$ will be
\begin{equation}\label{cur1}
{\bf J}({\bf r},t)= \delta({\bf r} - {\bf R}(t)){\bf v}(t).
\end{equation}

The macroscopic current averaged over time $\tau$ then reads
\begin{eqnarray}\label{cur2}
\nonumber
{\bf V}_\tau(t)&=&\tau^{-1}\int^{t+\tau}_t \int_{\bf r}{\bf J}({\bf r},t)\\
\nonumber
&=&\tau^{-1}\int^{t+\tau}_t \int_{\bf r} \delta({\bf r} - {\bf R}(t')){\bf v}(t') dt'\\
&=&\tau^{-1}\int^{t+\tau}_t {\bf v}(t') dt'.
\end{eqnarray}
Average of the Eq. \eqref{compacteomsup} over time duration $\tau$ gives
\begin{equation}\label{compacteomsup1}
\bm{\Pi} \cdot {\bf V}_\tau(t) = \bm{\mathcal{S}}_\tau(t) +\bm{\mathcal{F}}_{\tau}(t),
\end{equation}
where $\bm{\mathcal{S}}_\tau(t)=
\tau^{-1} \int^{t+\tau}_{t}{\bf S}(t')dt' $ and  $\bm{\mathcal{F}}_{\tau}(t)=\tau^{-1} \int^{t+\tau}_{t}{\bf f}(t')dt'$.
The probability density for $\bm{\mathcal{F}}_{\tau}(t)$ can be written as
\begin{eqnarray}\label{compacteomsup2}
\nonumber
P(\bm{\mathcal{F}}_{\tau}(t)=\textbf{A}) = \left\langle \delta \left(\frac{1}{\tau} \int^{t+\tau}_{t}{\bf f}(t')dt'-\textbf{A} \right)   \right\rangle_{\textbf{f}}\\
\nonumber
 =\dfrac{1}{(2 \pi)^d} \left\langle  \int_{\textbf{k}}\exp -i \left( \frac{1}{\tau} \int^{t+\tau}_{t}\textbf{k} \cdot {\bf f}(t')dt'-\textbf{k} \cdot \textbf{A} \right) \right\rangle _{\textbf{f}}\\
=\dfrac{1}{(2 \pi)^d}   \int_{\textbf{k}} \left\langle \exp -i  \frac{1}{\tau} \int^{t+\tau}_{t}\textbf{k} \cdot {\bf f}(t')dt' \right\rangle _{\textbf{f}} \exp i \textbf{k} \cdot \textbf{A}.
\end{eqnarray}
But from a well known identity \cite{v_bala_pramana_1993}
\begin{equation}\label{identity}
\left\langle \exp\left( \int_0^\tau H(t)g(t)dt\right) \right\rangle_{g}=\exp \left( \dfrac{1}{2} \int_0^\tau H(t)^{2}dt\right)
\end{equation}
 for any arbitrary function $H(t)$ and any white Gaussian noise $g(t)$ with zero mean and deviation one, 
\begin{eqnarray}\label{compacteomsup3}
\nonumber
P(\bm{\mathcal{F}}_{\tau}(t)=\textbf{A}) &=& 
\dfrac{1}{(2 \pi)^d}   \int_{\textbf{k}}  \exp \left(-\dfrac{1}{2 \tau^2} \int_0^\tau k^2 dt + i \textbf{k} \cdot \textbf{A} \right)\\
&=&\dfrac{1}{(2 \pi)^d}   \int_{\textbf{k}}  \exp \left(-\dfrac{k^2}{2 \tau} + i \textbf{k} \cdot \textbf{A} \right).
\end{eqnarray}
RHS of above equation is just Inverse Fourier Transform of a Gaussian function $ \exp-k^2 /2 \tau$, therefore the probability density for $\bm{\mathcal{F}}_{\tau}(t)$ becomes
\begin{eqnarray}\label{compacteomsup3}
P(\bm{\mathcal{F}}_{\tau}(t)=\textbf{A}) 
=\left(\dfrac{\tau}{2 \pi}\right)^{d/2}  \exp \left(-\dfrac{\tau A^2}{2} \right).
\end{eqnarray}
From Jacobian transformation the probability density for ${\bf V}_\tau(t)$ is
\begin{eqnarray}\label{compacteomsup4}
P({\bf V}_\tau(t))=\mbox{det} \left( \dfrac{\partial \bm{ \mathcal{F} }_\tau}{\partial {\bf V}_\tau}\right) P(\bm{\mathcal{F}}_{\tau}(t)).  
\end{eqnarray}
But from Eq. \eqref{compacteomsup1} 
\begin{equation}\label{22}
\dfrac{\partial \bm{ \mathcal{F} }_\tau}{\partial {\bf V}_\tau}= \bm{\Pi},
\end{equation}
hence 
\begin{eqnarray}\label{compacteomsup5}
\nonumber
P({\bf V}_\tau(t)&=&{\bf V})
=\mbox{det}\bm{\Pi} \left( \dfrac{\tau}{2 \pi}\right)^{d/2}\times\\ &&\exp \left(-\dfrac{\tau}{2} \left(\bm{\Pi} \cdot {\bf V}- \bm{\mathcal{S}}_\tau(t)\right)^2 \right).
\end{eqnarray}
Since $\bm{\Pi}^T \bm{\Pi}=(2\bsf{D})^{-1}$,
\begin{eqnarray}\label{33}
\nonumber
\mbox{det}\bm{\Pi}&=&(\mbox{det}({2\bsf{D}}^{-1}))^{1/2}\\
\nonumber
&=&2^{-d/2}({\mbox{det}{\bsf{D}})}^{-1/2}
\end{eqnarray}
and 
\begin{eqnarray}\label{compacteomsup6}
\nonumber
&&\left(\bm{\Pi} \cdot {\bf V} - \bm{\mathcal{S}}_\tau(t)\right)^2\\
\nonumber
&&=\left(\bm{\Pi} \cdot {\bf V}- \bm{\mathcal{S}}_\tau(t) \right)^T \left(\bm{\Pi} \cdot {\bf V}- \bm{\mathcal{S}}_\tau(t)\right)\\
\nonumber
&&=\left(  {\bf V}-\bm{\Pi}^{-1} \cdot \bm{\mathcal{S}}_\tau(t) \right)^T \bm{\Pi}^T \bm{\Pi} \left(  {\bf V}- \bm{\Pi}^{-1} \bm{\mathcal{S}}_\tau(t)\right)\\
&&=\frac{1}{2}\left(  {\bf V}-\bm{\Pi}^{-1} \cdot \bm{\mathcal{S}}_\tau(t) \right)^T {\bsf{D}}^{-1} \left(  {\bf V}- \bm{\Pi}^{-1} \bm{\mathcal{S}}_\tau(t)\right).
\end{eqnarray}
Eq. \eqref{compacteomsup5} then becomes
\begin{eqnarray}\label{compacteomsup7}
\nonumber
P_{\tau}(\textbf{V}_{\tau} = {\bf V})=
\left(\mbox{det}\bsf{D}\right)^{-1/2} \left({\tau \over 4 \pi}\right)^{d/2} \times\\
e^{- {\tau \over 4} ({\bf V} - \bm{\Pi}^{-1} \cdot \bm{\mathcal{S}}_\tau(t))^T \bsf{D}^{-1}({\bf V} - \bm{\Pi}^{-1} \cdot \bm{\mathcal{S}}_\tau(t))}.
\end{eqnarray}
If the average  velocity of the particle in steady state is $ {\bf v}_0$, 
the  $\bm{\mathcal{S}}_\tau(t)$ can be approximated by $\bm{\Pi} \cdot {\bf v}_0$, thus probability density for ${\bf V}_\tau(t)$ becomes 
\begin{eqnarray}\label{compacteomsup8}
\nonumber
P_{\tau}(\textbf{V}_{\tau} = {\bf V})=
\left(\mbox{det}\bsf{D}\right)^{-1/2} \left({\tau \over 4 \pi}\right)^{d/2}\times\\
e^{- {\tau \over 4} ({\bf V} -  {\bf v}_0)^T \bsf{D}^{-1}({\bf V} -  {\bf
v}_0)}.
\end{eqnarray}
\subsection{In $2$ dimensions}\label{apend2}
In two dimensions, for diagonal $\bsf{D}= \mbox{diag}(D_{\parallel}, D_{\perp})$, Eq. \eqref{compacteomsup8} becomes 
\begin{eqnarray}
\label{probVtau_avsub}
\nonumber
P_\tau(\textbf{V}_\tau &=& {\bf V})=\dfrac{\tau}{4 \pi \sqrt{D_{\parallel} D_{\perp}}
}\times\\
&&\exp
\left[ -\dfrac{\tau \left(V_{\parallel}-v_0^{\parallel}\right)^2}{4 D_{\parallel}}
-\dfrac{\tau \left(V_{\perp}-v_0^{\perp} \right)^2}{4 D_{\perp}}\right].
\end{eqnarray}
Two coarse-grained currents $V$ and $V'$ satisfying the condition
\begin{equation}\label{ifr1sub}
\dfrac{{V_{\parallel}}^2}{D_{\parallel}}+\dfrac{{V_{\perp}}^2}{D_{\perp}}=\dfrac{{V'_{\parallel}}^2}{
D_{\parallel}}+\dfrac{{V'_{\perp}}^2}{ D_{\perp}},
\end{equation}
i.e., which lie on the ellipse
${{V_{\parallel}}^2}/{D_{\parallel}}+{{V_{\perp}}^2}/{D_{\perp}}=\text{constant}$, obey
\begin{equation}\label{ifrsub}
\lim_{\tau \to \infty }\frac{1}{\tau} \ln
\dfrac{P_{\tau}(\textbf{V}_\tau = {\bf V})}{P_{\tau}(\textbf{V}_\tau =
{\bf V}')}=\bm{ \epsilon}
\cdot (\textbf{V}-\textbf{V}'). 
\end{equation}
For currents ${\bf V}$, ${\bf V}'$ with \textit{equal} magnitude $V$ the
result of \cite{PNASIFR} can be re-expressed as
\begin{eqnarray}\label{new1sub}
\nonumber
&&\lim_{\tau \to \infty }\frac{1}{\tau} \ln
\dfrac{P_{\tau}(\textbf{V}_\tau ={\bf V})}{P_{\tau}(\textbf{V}_\tau = {\bf
V}')}\\
&=& V \left[ \dfrac{v^{{\parallel}}_0}{2 D_{\parallel}} (\cos \theta - \cos
\theta')+\dfrac{v^{\perp}_0}{2 D_{\perp}} (\sin \theta - \sin \theta')\right]\nonumber
\\
&+&\dfrac{V^2
(D_{\parallel}-D_{\perp})}{4D_{\perp} D_{\parallel}}( \cos^2 \theta -\cos^2 \theta')
\end{eqnarray}
The true large-deviation function will presumably not have
the quadratic form
implied by \eqref{probVtau_avsub}. 
One should note that this calculation gives us a value 
\begin{equation}
\bm{ \epsilon}=({v^{\parallel}_0}/{2D_{\parallel}}, {v^{{\perp}}_0}/{2D_{\perp}}) 
\end{equation}
for $\bm{ \epsilon}$ in \eqref{ifrsub} in terms of independently measurable
quantities.


\begin{thebibliography}{10}
\bibitem{timerev} Although our system is not obviously reversible at the level
of the dynamics of rod and bead coordinates, with a one-way mean flux of
energy from the shaker into unresolved internal modes and ultimately into
radiant heat, the results of \cite{NK} prompt this further exploration.
\bibitem{footnote_nongaussian} \look{Perhaps this is because rolling motion, with some degree of persistence and hence fat tails, contributes to the lateral movement of the rod.}
\end{thebibliography}
\end{document}